\documentclass[sigconf]{acmart}
\usepackage{booktabs} 
\usepackage{spreadtab}
\usepackage{amsmath}
\usepackage{float}
\usepackage{graphicx}
\usepackage[skip=5pt]{caption}

\usepackage{color}
\usepackage{tcolorbox}
\usepackage{array}
\usepackage{listings}  
\usepackage{CJK}
\usepackage{hyperref}

\usepackage{adjustbox}
\usepackage{subcaption}
\DeclareCaptionSubType*[Alph]{table}
\DeclareCaptionLabelFormat{mystyle}{Table~\bothIfFirst{#1}{ ̃}#2}
\usepackage{multirow}

\begin{document}

\title{Context-Aware Sentence/Passage Term Importance Estimation \\ For First Stage Retrieval}

\author{Zhuyun Dai}
\affiliation{%
   \institution{Carnegie Mellon University}
   }
\email{zhuyund@cs.cmu.edu}

\author{Jamie Callan}
\affiliation{%
   \institution{Carnegie Mellon University}
   }
\email{callan@cs.cmu.edu}

\begin{abstract}

\end{abstract}




\begin{abstract}

Term frequency is a common method for identifying the importance of a term in a query or document. But it is a weak signal, especially when the frequency distribution is flat, such as in long queries or short documents where the text is of sentence/passage-length. This paper proposes a Deep Contextualized Term Weighting framework that learns to map BERT's contextualized text representations to context-aware term weights for sentences and passages.

When applied to passages, \texttt{DeepCT-Index} produces term weights that can be stored in an ordinary inverted index for passage retrieval.  When applied to query text, \texttt{DeepCT-Query} generates a weighted bag-of-words query.  Both types of term weight can be used directly by typical first-stage retrieval algorithms.  This is novel because most deep neural network based ranking models have higher computational costs, and thus are restricted to later-stage rankers.

Experiments on four datasets demonstrate that \texttt{DeepCT}'s deep contextualized text understanding greatly improves the accuracy of first-stage retrieval algorithms.
\end{abstract}


\maketitle

\section{Introduction}

State-of-the-art search engines use ranking pipelines in which an efficient \textit{first-stage} uses a query to fetch an initial set of documents from the document collection, and one or more \textit{re-ranking} algorithms improve and prune the ranking.  Typically the first stage ranker is a Boolean, probabilistic, or vector space bag-of-words retrieval model that fetches information from an inverted index.  One key characteristic of first-stage ranking algorithms is how they quantify the contribution of each query or document term.  Most retrieval methods use frequency-based signals such as document and query term frequency (\textit{tf}, \textit{qtf}) to determine the context-specific importance of a term, and and inverse document frequency (\textit{idf}) or a similar value to determine its importance globally.

Frequency-based term weights have been a huge success, but they are a crude tool. Term frequency does not necessarily indicate whether a term is important or central to the meaning of the text, especially when the frequency distribution is flat, such as in \emph{sentences and short passages}. Table \ref{tab:stomach-example} shows two passages from the MS MARCO machine reading comprehension dataset~\cite{nguyen2016ms}. If a user searches for `stomach', the two passages will be considered similarly relevant by frequency-based retrieval models because both mention the query word `stomach' twice; but the second passage is actually off-topic. To identify which terms are central requires a deeper understanding that considers the meaning of a word, the meaning of the entire text, and the role the word plays in the text. 

Recently, there has been
rapid progress in text understanding with the introduction of \emph{deep contextualized
word representations} such as ELMo~\cite{Peters:2018} and BERT~\cite{devlin2018bert}. 
These methods learn a neural language model from large text corpora. Each token is assigned a representation that is a function of not only the token itself, but also of the entire text. These neural language models have been shown to capture the characteristics of a word's semantics and syntax, and more importantly, how they vary across linguistic contexts~\cite{tenney2019you}.

\begin{table}[t]
\caption{Two passages that mention `\textbf{stomach}' twice.  Only the first passage is about the topic `stomach'. This paper proposes a method to weight terms by their roles in a specific text context, as shown by the heatmap over terms.}\label{tab:stomach-example}
\small
\begin{tabular}{p{0.96\columnwidth} }
\hline \hline

\begin{CJK*}{UTF8}{gbsn}
{\setlength{\fboxsep}{0pt}\colorbox{white!0}{\parbox{0.47\textwidth}{
\colorbox{red!0}{\strut In} \colorbox{red!0}{\strut some} \colorbox{red!0}{\strut cases,}  \colorbox{red!0}{\strut an} \colorbox{red!31}{\strut upset} \colorbox{red!32}{\strut \textbf{stomach}} \colorbox{red!0}{\strut is} \colorbox{red!0}{\strut the} \colorbox{red!0}{\strut result} \colorbox{red!0}{\strut of} \colorbox{red!0}{\strut an} \colorbox{red!1}{\strut allergic} \colorbox{red!1}{\strut reaction} \colorbox{red!0}{\strut to} \colorbox{red!0}{\strut a} \colorbox{red!0}{\strut certain} \colorbox{red!0}{\strut type} \colorbox{red!0}{\strut of} \colorbox{red!2}{\strut food}\colorbox{red!0}{\strut .} \colorbox{red!0}{\strut It} \colorbox{red!0}{\strut also} \colorbox{red!0}{\strut may} \colorbox{red!0}{\strut be} \colorbox{red!30}{\strut caused} \colorbox{red!0}{\strut by} \colorbox{red!0}{\strut an} \colorbox{red!0}{\strut irritation}\colorbox{red!0}{\strut .} \colorbox{red!0}{\strut Sometimes} \colorbox{red!0}{\strut this} \colorbox{red!0}{\strut happens} \colorbox{red!0}{\strut from} \colorbox{red!0}{\strut consuming} \colorbox{red!0}{\strut too} \colorbox{red!0}{\strut much} \colorbox{red!0}{\strut alcohol} \colorbox{red!0}{\strut or} \colorbox{red!0}{\strut caffeine}\colorbox{red!0}{\strut .} \colorbox{red!2}{\strut Eating} \colorbox{red!0}{\strut too} \colorbox{red!0}{\strut many} \colorbox{red!0}{\strut fatty} \colorbox{red!2}{\strut foods}  \colorbox{red!0}{\strut or} \colorbox{red!0}{\strut too} \colorbox{red!0}{\strut much} \colorbox{red!2}{\strut food} \colorbox{red!0}{\strut in} \colorbox{red!0}{\strut general} \colorbox{red!0}{\strut may} \colorbox{red!0}{\strut also} \colorbox{red!31}{\strut cause} \colorbox{red!0}{\strut an} \colorbox{red!31}{\strut upset} \colorbox{red!32}{\strut \textbf{stomach}}\colorbox{red!0}{\strut .} 
}}}
\end{CJK*} \\ \hline 

\begin{CJK*}{UTF8}{gbsn}
{\setlength{\fboxsep}{0pt}\colorbox{white!0}{\parbox{0.47\textwidth}{
\colorbox{red!0}{\strut All} \colorbox{red!12}{\strut parts} \colorbox{red!0}{\strut of} \colorbox{red!0}{\strut the} \colorbox{red!18}{\strut body} \colorbox{red!0}{\strut (}\colorbox{red!4}{\strut muscles} \colorbox{red!0}{\strut ,} \colorbox{red!4}{\strut brain}\colorbox{red!0}{\strut ,} \colorbox{red!3}{\strut heart}\colorbox{red!0}{\strut ,} \colorbox{red!0}{\strut and} \colorbox{red!2}{\strut liver}\colorbox{red!0}{\strut )} \colorbox{red!3}{\strut need} \colorbox{red!11}{\strut energy} \colorbox{red!0}{\strut to} \colorbox{red!7}{\strut work}\colorbox{red!0}{\strut .} \colorbox{red!0}{\strut This} \colorbox{red!11}{\strut energy} \colorbox{red!4}{\strut comes} \colorbox{red!0}{\strut from} \colorbox{red!0}{\strut the} \colorbox{red!6}{\strut food} \colorbox{red!0}{\strut we} \colorbox{red!6}{\strut eat}\colorbox{red!0}{\strut .} \colorbox{red!0}{\strut Our} \colorbox{red!16}{\strut bodies} \colorbox{red!12}{\strut digest} \colorbox{red!0}{\strut the} \colorbox{red!6}{\strut food} \colorbox{red!0}{\strut we} \colorbox{red!6}{\strut eat} \colorbox{red!0}{\strut by} \colorbox{red!2}{\strut mixing} \colorbox{red!0}{\strut it} \colorbox{red!0}{\strut with} \colorbox{red!2}{\strut fluids}\colorbox{red!0}{\strut (} \colorbox{red!2}{\strut acids} \colorbox{red!0}{\strut and} \colorbox{red!3}{\strut enzymes}\colorbox{red!0}{\strut )} \colorbox{red!0}{\strut in} \colorbox{red!0}{\strut the} \colorbox{red!8}{\strut \textbf{stomach}}\colorbox{red!0}{\strut .} \colorbox{red!0}{\strut When} \colorbox{red!0}{\strut the} \colorbox{red!8}{\strut \textbf{stomach}} \colorbox{red!9}{\strut digests} \colorbox{red!6}{\strut food}\colorbox{red!0}{\strut ,} \colorbox{red!0}{\strut the} \colorbox{red!3}{\strut carbohydrate} \colorbox{red!0}{\strut (}\colorbox{red!2}{\strut sugars} \colorbox{red!0}{\strut and} \colorbox{red!2}{\strut starches}\colorbox{red!0}{\strut )} \colorbox{red!0}{\strut in} \colorbox{red!0}{\strut the} \colorbox{red!6}{\strut food} \colorbox{red!2}{\strut breaks} \colorbox{red!0}{\strut down} \colorbox{red!0}{\strut into} \colorbox{red!0}{\strut another} \colorbox{red!3}{\strut type} \colorbox{red!0}{\strut of} \colorbox{red!4}{\strut sugar}\colorbox{red!0}{\strut ,} \colorbox{red!3}{\strut called} \colorbox{red!3}{\strut glucose}\colorbox{red!0}{\strut .}
}}}
\end{CJK*}\\ \hline \hline
\end{tabular}
\end{table}

This paper shows how to improve first-stage retrieval models using the contextualized word representations generated by BERT~\cite{devlin2018bert}. 
We present the Deep Contextualized Term Weighting framework (\texttt{DeepCT}). \texttt{DeepCT} is trained in a supervised manner to learn a BERT-based contextualized word representation model and a mapping function from representations to term weights. 
As a word's representation depends on its specific context, the estimated importance for the same term varies with the context. \texttt{DeepCT} can take up to 512 text tokens~\footnote{This is limited by the input length limitation of BERT~\cite{devlin2018bert}}. It can makes use of the linguistic context in sentence/passage-long text to generate more representative term weights, helping the cases where the original term frequency distribution is flat, such as retrieving passages or searching long queries.

One use of \texttt{DeepCT} is to identify \textbf{essential passage terms for passage retrieval}. As shown in Table~\ref{tab:stomach-example}, passages usually have flat term weight distribution, making term-frequency based retrieval less effective. We develop a novel \texttt{DeepCT-Index} that offline weights and indexes terms in passage-long documents.  It trains a \texttt{DeepCT} model to predict whether a passage term is likely to appear in relevant queries. 
The trained model is applied to every passage in the collection. This inference step is query-independent, allowing it to be done offline during indexing. The context-based passage term weights are scaled to \textit{tf}-like integers that are stored in an ordinary inverted index that can be searched efficiently by common first-stage retrieval models.

Another use of \texttt{DeepCT} is to identify \textbf{important query terms in long queries}. For long queries that mention many terms and concepts, it is important to identify which are central. We follow a query term weighting framework proposed by ~\citet{zheng2015learning} to develop \texttt{DeepCT-Query}. It trains \texttt{DeepCT} with signals from relevant query-document pairs,  weighting query terms by their possibilities to be mentioned by relevant documents. 
The predictions are used to generate weighted queries that can be used with widely-used retrieval models such as BM25~\cite{bm25} and query likelihood~\cite{Lavrenko2001RelevanceBasedLM}.

Our experiments demonstrate that \texttt{DeepCT} generates effective representations for both passages and queries that lead to large improvements in first-stage retrieval accuracy. 
Analysis shows that \texttt{DeepCT}'s main advantage is its ability 
to estimate term importance using the meaning of the context rather than term frequency signals, allowing the retrieval model to differentiate between key terms and other frequently mentioned but non-central terms.

Most of the prior neural-IR research~\cite{DRMM, dai2018convolutional}, including recent research on leveraging BERT for IR~\cite{nogueira2019passage, dai2019deeper}, focused on re-ranking stages due to the complexity of neural models. 
Our work adds the ability to improve existing first-stage rankers. 
More accurate first-stage document rankings provide better candidates for downstream re-ranking, which improves end-to-end accuracy and/or efficiency.

Section \ref{section:related-work} discusses related work. Section~\ref{section:method} describes the Deep Contextualized Term Weighting framework (\texttt{DeepCT}), its use for passage indexing (\texttt{DeepCT-Index}), and its use for for query weighting (\texttt{DeepCT-Query}).  
Sections \ref{section:exp-res-doc}-\ref{section:exp-set-qry} describe our methodologies and experiments. 
Section~\ref{section:conclusion} concludes \footnote{We have released software and data: \url{https://github.com/AdeDZY/DeepCT}}.

\section{Related Work}\label{section:related-work}

Two types of work are related: term weighting, and neural approaches for early-stage ranking.

\textbf{Term Weighting.} Bag-of-words retrieval models such as BM25 \cite{bm25} and query likelihood ~\cite{metzler2005markov} are the foundation of modern search engines due to their efficiency and effectiveness.  Most retrieval models use frequency-based signals such as \textit{tf}, \textit{ctf}, and \textit{df} to estimate how well each term represents a text (query, document).  A rich research literature studies how to better estimate term importance (e.g.,~\cite{mihalcea2004textrank, blanco2012graph, rousseau2013graph, zheng2015learning, bendersky2011parameterized, bendersky2010learning}), however frequency-based signals continue to dominate in both industry and academia.

For document term weighting, the most widely-used alternatives to frequency-based models are graph-based methods~\cite{mihalcea2004textrank, blanco2012graph, rousseau2013graph}. Graph-based approaches build a document graph, where the nodes represent terms and edges represent term co-occurrence within a maximum distance. Terms are ranked by graph ranking algorithms such as PageRank, and scores are indexed for retrieval~\cite{blanco2012graph, rousseau2013graph}.

Recently, a few neural IR approaches study document term weight from word embeddings~\cite{WeakNeuIR}, because word embeddings have the ability to encode some aspects of word meanings that are useful for determining word importance. They add a term-gating sub-network into neural re-rankers that learn global \textit{idf}-like term weights from word embeddings. It is not clear whether they can improve the initial ranking stage.

For query term weighting, one line of research focuses on feature-based approaches~\cite{bendersky2012effective, bendersky2011parameterized, bendersky2010learning}. They require running the query against a document collection to generate features. These methods improve search accuracy compared to frequency-based query term weighting, but the use of pseudo-relevant feedback causes extra computational cost. To predict query term weights from just the text content, ~\citet{zheng2015learning} proposed a word-embedding based method called \texttt{DeepTR}.  \texttt{DeepTR} constructs a feature vector for each query term using the difference between the term's word2vec~\cite{word2vec} embedding and the average query embedding. It then learns a regression model to map the feature vector onto the term's ground truth weight (term recall weight~\cite{zhao2010term}). The estimated weights are used to generate bag-of-words queries that can be searched in first-stage retrieval. Later, ~\citet{DRMM} used a similar word-embedding based query term weighting method in neural re-rankers.

As discussed above, in both document and query term weighting problems, word embeddings have gained increasing attention due to their ability to encode certain aspects of text meaning~\cite{WeakNeuIR, zheng2015learning, DRMM}.
However, word embedding based approaches face a critical issue in capturing the context, as classic word embeddings are context-independent -- a word always has the same embedding regardless of the text context. Recently, contextualized neural language models have been proposed to model the linguistic context~\cite{Peters:2018, devlin2018bert}. In contextualized neural language models, a word's representation is a function of the entire text input. These representations were shown to capture 
a word's certain syntactic features and semantic features in the text content~\cite{tenney2019you}. Currently, the best performing neural language model is BERT~\cite{devlin2018bert}. BERT has received a lot of attention for IR, mainly focused on using it as a black-box re-ranking model to predict query-document relevance scores~\cite{dai2019deeper, nogueira2019passage}. This work attempts to leverage the BERT's advantages to explicitly model query/document term importance for efficient first-stage retrieval.

\textbf{Neural Approaches for Early Stage Ranking}. Most neural ranking models use continuous representations that enable a query to match every document to some degree, which makes them impractical for first-stage rankers.  Recent research addresses this efficiency problem in two ways, both using latent representations.  One approach learns low dimensional dense embeddings of queries and documents that support fast retrieval using approximate nearest neighbor search in the embedding space~\cite{aumuller2017distance}, however there is an efficiency vs.~accuray tradeoff \cite{boytsov-cikm16}. 
A second approach is to learn high-dimensional but sparse latent representations in which queries and documents are represented by a set of `latent words'~\cite{salakhutdinov2009semantic,zamani2018neuralre}. Sparse `latent words' allow inverted index style look up, which is efficient, but also introduces the specificity vs. exhaustiveness tradeoff found in all controlled vocabularies \cite{salton83}.

Recently, ~\citet{nogueira2019doc2qry} proposed \texttt{Doc2Query} that uses neural models to modify the discrete word-based document representations. It generates potential queries from documents using neural machine translation and indexes those queries as document expansion terms. We are not aware of other first-stage neural models that use the discrete word-based text representations and large vocabularies that have been fundamental to modern search engines.

\section{Deep Contextualized Document and Query Term Weighting}\label{section:method}
This section presents the Deep Contextualized Term Weighting framework (\texttt{DeepCT}), and how it is used to weight query terms (\texttt{DeepCT-Query}) and index documents (\texttt{DeepCT-Index}).

\subsection{DeepCT Framework}\label{subsection: DeepCT}


\texttt{DeepCT} includes two main components: generating contextualized word embeddings through BERT, and predicting term weights through linear regression. 

\textbf{Contextualized word embedding generation}. To estimate the importance of a word in a specific text, the most critical problem is to generate features that  characterize a word's relationships to the text context. Recent contextualized neural language models like ELMo~\cite{Peters:2018} and BERT~\cite{devlin2018bert} have been shown to capture such properties through a deep neural network. \texttt{DeepCT} leverages one of the best performing models, BERT, to extract a word's context features. BERT uses an attention mechanism where a word gradually absorbs context information based on its attention to every other word in the same text. \citet{devlin2018bert} provide more detail. 

\textbf{Map to target weights}. The contextualized word embedding is a feature vector that characterizes the word's syntactic and semantic role in a given context. \texttt{DeepCT} linearly combines the features into a word importance score:
\begin{equation}
    \hat{y}_{t,c} = \vec{w} T_{t,c} + b
    \label{eq:regression}
\end{equation}
where $T_{t,c}$ is token $t$'s contextualized embedding in text $c$; and, $\vec{w}$ and $b$ are the linear combination weights and bias. 

\texttt{DeepCT} is trained with a per-token regression task. Given the ground truth term weight for every word in text $c$, denoted as $\{y_{1,c}, \ldots, y_{N,c}\}$, \texttt{DeepCT} aims to minimize the mean square error (MSE) between the predicted weights $\hat{y}$ and the target weights $y$:

\begin{equation}
    loss_{MSE} = \sum_c\sum_t(y_{t,c} - \hat{y}_{t,c})^2.
    \label{eq:mse}
\end{equation}

The range of possible predicted term weights $\hat{y}_{t,c}$ is $(-\infty, \infty)$, but in practice most predictions fall into $[0, 1]$ because, as shown in Sections~\ref{subsection: DeepCT-query} and ~\ref{subsection: DeepCT-query}, the ground truth weights are $[0,1]$. 
Our query/document weighting approaches accept any non-negative weights; terms with negative weights are discarded.

The tokenization step of BERT generates subwords for unseen words (e.g. "DeepCT" is tokenized into "deep" and "\#\#ct"). We use the weight for the first subword as the weight of the entire word; other subwords are masked out when computing the MSE loss.

The \texttt{DeepCT} model, from BERT to the regression layer, is optimized end-to-end. The BERT component is initialized with a pre-trained BERT model to reduce over-fitting. It is fine-tuned to align the contextualized word embeddings with the term-prediction task. The last regression layer is learned from scratch.  

\texttt{DeepCT} is a general framework that learns term weights with respect to a linguistic context. \texttt{DeepCT} can \textit{learn} different definitions of term importance depending on how ground truth term weights are defined. The predicted term weights can also be \textit{used} differently depending on the task. Below, we describe two approaches to using the \texttt{DeepCT} framework to improve first-stage retrieval.

\subsection{Index Passages with DeepCT}\label{subsection: DeepCT-index}

A novel use of \texttt{DeepCT} is to identify terms that are central to the meaning of a passage, or a passage-long document, for efficient and effective passage/short-document retrieval. In this paper, We will refer to both passages in passage retrieval and passage-long documents in short-document retrieval as `passages'.

As shown in the `stomach' example in Table~\ref{tab:stomach-example},   classic term frequency signals cannot tell whether the text is centered around a term or just mentions that term when discussing some other topic. 
This issue is especially difficult in first-stage full-collection ranking, where complex features and models are too expensive to apply. \texttt{DeepCT-Index} uses \texttt{DeepCT} to weight passage terms and stores the weights in a typical inverted index.

\textbf{Target Term Weights for Training \texttt{DeepCT}.}  
Proper target term weights should reflect whether a term is essential to the passage or not.
We propose \emph{query term recall} as an estimation of the ground truth passage term importance: 
\begin{equation}
    QTR(t,d) = \frac{|Q_{d,t}|}{|Q_d|}.
\end{equation}
$Q_d$ is the set of queries for which passage,= $d$ is judged relevant. $Q_{d, t}$ is the subset of $Q_d$ that contains term $t$, and $QTR(t,d)$ is the query term recall weight for $t$ in $d$. $QTR$ is in the range of $[0, 1]$. Query term recall is based on the assumption that search queries can reflect the key idea of a document. Words that appear in relevant queries are more important than other words in the document. 

The training requires relevant query-passage pairs. The model takes the text content of a passage, make predictions, and compute the loss with target weights generated from its related queries. During inference, the model needs only the passage. 

\textbf{Index with predicted term weights.}
Once \texttt{DeepCT} learns model parameters, it can make estimates for any passage without the need of queries. This allows estimated term weights to be calculated and stored during offline indexing. The index can be searched efficiently by online services.

We apply the trained \texttt{DeepCT} model to all passages in the collection. 
The predicted weights are scaled from a [0..1] range to an integer range that can be used with existing retrieval models. We call this weight $\text{TF}_{\text{DeepCT}}$ to convey that it is an alternate way of representing the importance of t erm $t$ in passage $d$: 
\begin{equation}\label{eq:tf}
   \text{TF}_{\text{DeepCT}}(t, d) = round(\hat{y}_{t,d} * N),
\end{equation}
where $\hat{y}_{t,d}$ is the predicted term weights for term $t$ in passage $d$. $N$ scales the predicted weights into a integer range. We use $N=100$ in this work as two digits of precision is sufficient for this task.

$\text{TF}_{\text{DeepCT}}$ is used to \emph{replace} the original TF field in the inverted index. The postings of term $t$ changes from [docid(d), $\text{TF}(t,d)$] into [docid(d), $\text{TF}_{\text{DeepCT}}(t,d)$]. The new index, \texttt{DeepCT-Index}, can be searched by  mainstream bag-of-words retrieval model like BM25 or query likelihood model (QL). The context-based term weight $\text{TF}_{\text{DeepCT}}$ is expect to bias the retrieval models to central terms in the pasessag, preventing off-topic passages being retrieved.

\textbf{Efficiency}. \texttt{DeepCT-Index} does not add latency to the search system. 
The main difference between \texttt{DeepCT-Index} and a typical inverted index is that the term importance weight is based on $\text{TF}_\text{DeepCT}$ instead of TF.  This calculation is done offline.  No new posting lists are created, thus the query latency does not become longer. To the contrary, a side-effect of Eq~\ref{eq:tf} is that $\text{TF}_\text{DeepCT}$ of some terms becomes $0$.  This may be viewed as a form of index pruning~\cite{carmel2001static}.  We leave that aspect of this work for future investigation.

\subsection{Query Term Weighting with DeepCT}\label{subsection: DeepCT-query}
Another straightforward use of \texttt{DeepCT} for IR is to weight query terms in long queries. For long queries that mention many terms and concepts, it is important to identify which are central. For example, given the query \textit{``Find locations of volcanic activity which occurred within the present day boundaries of the U.S and its territories''}, an ideal system would understand that \textit{``volcanic activity''} is the key concept, and that \textit{``boundaries of the U.S''} maybe not be necessary in some corpora.  ~\citet{zheng2015learning} proposed a word2vec-based query term re-weighting framework, called \texttt{DeepTR}, that efficiently re-weights bag-of-words queries. We follow the \texttt{DeepTR} framework, but replace the word2vec-based model with \texttt{DeepCT}. We call the proposed approach \texttt{DeepCT-Query}.

\textbf{Target Term Weights for Training \texttt{DeepCT}.} 
Inspired by ~\citet{zheng2015learning}, \texttt{DeepCT-Query} uses \emph{term recall}~\cite{zhao2010term}:
\begin{equation}
    TR(t,q) = \frac{|D_{q,t}|}{|D_q|}.
\end{equation}
$D_q$ is the set of documents that are relevant to the query.  $D_{q, t}$ is the subset of relevant documents that contains term $t$. Their ratio, $TR(t,q)$, is the term recall weight for term $t$ in query $q$.  Term recall is in the range of $[0, 1]$. Term recall is based on the assumption that a query term is more important if it is mentioned by more relevant documents. 

Training requires relevant query-document pairs. The model takes the text content of a query, makes predictions, and computes the loss with target weights generated from the relevant documents. During inference, the model only needs the query. 

\textbf{Re-weight queries with predicted term weights.} When a query is received (\textit{online}), the trained model is used to predict importance weights for each term. Following \texttt{DeepTR}~\cite{zheng2015learning}, we use the estimated query term weights to generate bag-of-words  queries (\texttt{BOW}) and sequential dependency model queries (\texttt{SDM}). For example, in the Indri~\cite{Strohman05Indri} query language, the original \texttt{BOW} query \texttt{"apple pie"} is re-formulated into \texttt{\#weight(0.8 apple 0.7 pie)}\footnote{\texttt{\#weight} is Indri's probabilistic weighted-AND operator.} for predicted weights \texttt{apple:0.8}, \texttt{pie:0.7}. Sequential dependency model adds bigrams and word co-occurrences within a window to the query. We use the re-weighted \texttt{BOW} query to replace the bag-of-words part of the \texttt{SDM} query. Terms with non-positive weights are discarded.  
In terms of efficiency,  predicting term weights for a new query is simply feeding-forward the query string through \texttt{DeepCT}. We use \texttt{Bow-DeepCT-Query} and \texttt{SDM-DeepCT-Query} to denote the re-weighted bag-of-words and sequential dependency queries.

\section{Experimental Methodology for D\texorpdfstring {\lowercase{eep}}{eep}CT-I\texorpdfstring{\lowercase{ndex}}{ndex}}
\label{section:exp-set-doc}

This section presents the experimental methodology for the second task --  passage term weighting and indexing. 

\textbf{Datasets}: The current implementation of BERT supports texts of up to 512 tokens, thus we selected two collections that consist primarily of passages: \textbf{MS MARCO}~\cite{nguyen2016ms} and \textbf{TREC-CAR}~\cite{dietz2017trec}. Passages and short documents are also the case where weighting by term frequency is less effective, because the \textit{tf} distribution tends to be flat.
The experimental settings mainly follow the methodology used in previous neural passage ranking/re-ranking studies ~\cite{nogueira2019passage, nogueira2019doc2qry}

MS MARCO~\cite{nguyen2016ms} is a question-to-passage retrieval dataset with $8.8$M passages. Average passage length is around $55$ words. The training set contains approximately
$0.5$M pairs of queries and relevant passages. On
average each query has one relevant passage. The
development (dev) set contains 6,980 queries and their relevance labels.
The test set contains 6,900 queries, but the relevance labels are hidden by Microsoft. Therefore, \emph{the dev set is our main evaluation set.} 
In a few experiments, we also evaluated on the test set by submitting our rankings to the MS MARCO competition.

TREC-CAR~\cite{dietz2017trec} consists of $29.7$M English
Wikipedia passages with an average length of $61$ words. Queries and relevant passages are  generated synthetically. A query is the concatenation of a Wikipedia article title with the title of one of its sections. Following prior work~\cite{nogueira2019passage,nogueira2019doc2qry}, we use the automatic relevance judgments, which treats paragraphs within the section as relevant to the query. The training set and validation set have $3.3$M query-passage pairs and $0.8$M pairs respectively. The test query set contains 1,860 queries with an average of $2.5$ relevant paragraphs per query.

\textbf{Baselines}: Experiments were done with done with three baseline and three experimental indexing methods, as described below.

\texttt{tf} index is a standard \textit{tf}-based index, e.g., as used by BM25.

\texttt{TextRank}~\cite{mihalcea2004textrank} is a widely-used unsupervised graph-based term weighting approach. We use the open source PyTextRank implementation\footnote{\url{https://github.com/DerwenAI/pytextrank}}. Term weights from TextRank are in the range (0, 1);  we scale them to integers as described in Eq.~\ref{eq:tf} for indexing. 


\texttt{Doc2Query}~\cite{nogueira2019doc2qry} is a supervised neural baseline. It trains a neural sequence-to-sequence model to generate potential queries from passages, and indexes the queries as document expansion terms.  \texttt{Doc2Query} implicitly re-weights terms because important passage terms are likely to appear in the generated queries. We use the \texttt{Doc2Query} MS MARCO index released by the authors.  No such index is available for TREC-CAR, so we use published values for that dataset~\cite{nogueira2019doc2qry}. \texttt{Doc2Query} was the best-performing first-stage ranking method on MS MARCO at the time this paper was written.

The three experimental indexing methods include the proposed \texttt{DeepCT-Index} and two variants using different embeddings.

$\texttt{DeepCT}_\texttt{W}$-\texttt{Index} replaces the BERT component in \texttt{DeepCT} with context-independent word embeddings. To provide context to each word, we modeled the passage using the average word embeddings and subtracted the passage embedding from each word's embedding, which was inspired by~\cite{zheng2015learning}. The embeddings are initialized by word2vec~\cite{word2vec} and fine-tuned during training.

$\texttt{DeepCT}_\texttt{E}$-\texttt{Index} replaces the BERT component in \texttt{DeepCT} with ELMo~\cite{Peters:2018}. ELMo is a pre-trained, context-aware text representation model that differs from BERT in the network architecture and pre-training task. ELMo was initialized with the pre-trained model described by~\citet{Peters:2018} and fine-tuned during training.

\textbf{Indexing, Retrieval, and Evaluation}: 
First-stage ranking was done by two popular retrieval models: \texttt{BM25} and query likelihood with Jelinek-Mercer smoothing (\texttt{QL}). We used the Anserini toolkit implementations. We fine-tuned \texttt{BM25} parameters $k_1$ and $b$, and \texttt{QL} smoothing factor $\lambda$ through a parameter sweep on $500$ queries from the training set.
Re-ranking was done by two re-rankers: Conv-KNRM~\cite{dai2018convolutional} and a BERT Re-ranker~\cite{nogueira2019passage}. We used the trained re-rankers provided by the authors, and applied them to re-rank up to a $1000$ passages from a first-stage ranking.  

Ranking/re-ranking results were evaluated by Mean Reciprocal Rank at 10 passages (MRR@10), the official MS MARCO evaluation metric. For TREC-CAR, we also report MAP at depth 1,000 following the evaluation methodology used in previous work~\cite{nogueira2019passage,nogueira2019doc2qry}.

\textbf{DeepCT Settings}: 
The BERT part of \texttt{DeepCT-Index} was initialized with pre-trained parameters. For MS MARCO, we used the official pre-trained BERT (uncased, base model)~\cite{devlin2018bert}. TREC-CAR cannot use the official model because its testing documents overlapped with BERT's pre-training documents.  We used a BERT model from~\citet{nogueira2019passage} where the overlapping documents are removed from the pre-training data. 
After initialization, \texttt{DeepCT} is trained for $3$ epochs on the training split of our datasets, using a learning rate of $2e^{-5}$ and a max input text length of $128$ tokens.

\begin{table*}[t]
\caption{Ranking accuracy of \texttt{BM25} and \texttt{QL} using indexes built with three baselines and three \texttt{DeepCT-Index} methods. Win/Tie/Loss are the number of queries improved, unchanged, or hurt, compared to \texttt{tf} index on MRR@10. $^*$ and $^\dagger$ indicates statistically significant improvements over \texttt{tf} index and \texttt{Doc2Query}. \texttt{Doc2Query} results for TREC-CAR are from~\citet{nogueira2019doc2qry}; statistically significance for these results are unknown.}
\label{tab:retrieval}
\setlength{\tabcolsep}{3pt}
\renewcommand{\arraystretch}{1}

\begin{tabular}{l|lc|lc||llc|llc}
\hline \hline
\multirow{3}{*}{Index} & \multicolumn{4}{c||}{MS MARCO dev}                   & \multicolumn{6}{c}{TREC-CAR}                              \\ \cline{2-11}
                       & \multicolumn{2}{c|}{\texttt{BM25}} & \multicolumn{2}{c||}{\texttt{QL}}  & \multicolumn{3}{c|}{\texttt{BM25}}      & \multicolumn{3}{c}{\texttt{QL}}    \\ 
                       & {\small MRR@10}  & {\small W/T/L}          & {\small MRR@10} & {\small W/T/L}          & {\small MRR@10} & {\small MAP}   & {\small W/T/L}        & {\small MRR@10} & {\small MAP}   & {\small W/T/L}    \\ \hline
{\small \texttt{tf index}}               & 0.191   & -/-/-          & 0.189  & -/-/-          & 0.233  & 0.174 & -/-/-        & 0.211  & 0.162 & -/-/-    \\ \hline
{\small \texttt{TextRank}}               & 0.130   & 662/4556/1762  & 0.134  & 702/4551/1727  & 0.160  & 0.120 & 167/1252/441 & 0.157  & 0.118 &  166/1327/367        \\
{\small \texttt{Doc2Query}}              & 0.221$^*$   & 1523/4431/1026 & 0.224$^*$  & 1603/4420/957  & -      & 0.178 & -/-/-       & -       & -    & -/-/- \\ \hline
{\small \texttt{DeepCT-Index}}           & \textbf{0.243}$^{*\dagger}$    & 2022/3861/1097 & \textbf{0.230}$^*$  & 1843/4027/1110 & \textbf{0.332}$^*$  & \textbf{0.246}$^*$ & 615/1035/210 & \textbf{0.330}$^*$  & \textbf{0.247}$^*$ & 645/1071/144         \\
{\small $\texttt{DeepCT}_\texttt{W}$\texttt{-Index} }         & 0.174   & 931/4804/1245  & 0.168  & 867/4793/1320  & 0.205  & 0.147 & 250/1311/309 & 0.192  & 0.139 &  245/1345/280        \\
{\small $\texttt{DeepCT}_\texttt{E}$\texttt{-Index} }          &   0.234$^{*\dagger}$      &     1891/4139/950           &   0.220$^*$     &    1726/4210/1044    &    0.280$^*$    &   0.201$^*$    &    516/1144/200         &     0.276$^*$   &   0.197$^*$    &    540/1190/130         
\\ \hline \hline
\end{tabular}
\end{table*}

\section{D\texorpdfstring{\lowercase{eep}}{eep}CT-I\texorpdfstring{\lowercase{ndex}}{ndex} Results}

\label{section:exp-res-doc}
The next two subsections describe experiments that investigated first-stage search accuracy using \texttt{DeepCT-Index} indexes, and why \texttt{DeepCT\-Index} term weights are effective. 

\subsection{Retrieval Accuracy of DeepCT-Index}

This section examines whether \texttt{DeepCT-Index} improves first-stage retrieval accuracy over baseline term weighting methods. Then it compares the first-stage retrieval accuracy of \texttt{DeepCT-Index} to several single-/multi-stage competition search systems. Finally, it studies the impact of a first-stage ranker using \texttt{DeepCT-Index} on the end-to-end accuracy of multi-stage search systems.

\textbf{DeepCT-Index Retrieval Performance}.
Table~\ref{tab:retrieval} shows the first-stage retrieval accuracy of \texttt{BM25} and \texttt{QL} using indexes generated by six methods.  The \texttt{TextRank} index failed to beat the common \texttt{tf} index. The \texttt{Doc2Query} index was effective for MS MARCO, but only marginally better for TREC-CAR.    

\texttt{DeepCT-Index} outperformed the baselines by large margins. It improved \texttt{BM25} by $27\%$ on MS MARCO and $46\%$ on TREC-CAR. It produced similar gains for \texttt{QL}, showing that \texttt{DeepCT-Index} is useful to different retrieval models. Win/Loss analysis shows that \texttt{DeepCT-Index} improved 25-35\% of the queries and hurt 8-16\%.  

\texttt{DeepCT-Index} also surpassed the neural baseline \texttt{Doc2Query} by large margins. \texttt{DeepCT-Index} and \texttt{Doc2Query} were trained using the same data, demonstrating the advantage of \texttt{DeepCT-Index}'s explicit term-weighting  over \texttt{Doc2Query}'s implicit term-weighting through passage expansion.

Results for $\texttt{DeepCT}_\texttt{W}$\texttt{-Index} and $\texttt{DeepCT}_\texttt{E}$\texttt{-Index} demonstrate the importance of context.
Non-contextual word2vec embeddings produced term weights that were less effective than \textit{tf}.  ELMo produced more effective term weights, but BERT's stronger use of context produced the most effective weights.  These results also show the generality of the \texttt{DeepCT} framework. 

\begin{table}[t]
\caption{Retrieval accuracy of BM25 with \texttt{DeepCT-Index}  compared with several representative ranking/re-ranking systems on the MS MARCO dataset using official evaluation on dev and hidden test set.\protect\footnotemark}\label{tab:leaderboard}

\begingroup
\setlength{\tabcolsep}{4pt}

\begin{tabular}{c|l|ll|ll}
\hline \hline

\multicolumn{1}{l}{}                                                            &   Ranking    Method                                                                          & \multicolumn{2}{c|}{\begin{tabular}[c]{@{}c@{}}dev\\ MRR@10\end{tabular}} & \multicolumn{2}{c}{\begin{tabular}[c]{@{}c@{}}test\\ MRR@10\end{tabular}} \\ \hline
\multirow{3}{*}{\begin{tabular}[c]{@{}c@{}}
{\small Single-}\\{\small Stage}\end{tabular}}   & {\small Official BM25}                                                      & 0.167                              & -30\%                             & 0.165                              & -31\%                              \\

                                                                                & {\small \texttt{Doc2Query BM25}~\cite{nogueira2019doc2qry}    }                       & 0.215                              & -12\%                             & 0.218                              & -9\%                               \\ 
                                                                                
                                                                                 &  {\small \texttt{DeepCT-Index BM25} }& 0.243                              & --                                  & 0.239                              & --       \\ 
                                                                                                             \hline \hline
  \multirow{5}{*}{\begin{tabular}[c]{@{}c@{}} {\small Multi-}\\{\small Stage}\end{tabular}}                                                                              & {\small Feature-based LeToR } & 0.195 & -20\% & 0.191 & -20\% \\
 & {\small K-NRM~\cite{K-NRM}   }                                                                         & 0.218                              & -10\%                             & 0.198                              & -17\%                              \\  \cline{2-6}
                                                                                &  {\small Duet V2~\cite{mitra2019updated}}                                                & 0.243                              & +0\%                              & 0.245                              & +2\%                               \\
                                                                                & {\small Conv-KNRM~\cite{dai2018convolutional}           }                                                            & 0.247                              & +2\%                              & 0.247                              & +3\%                               \\  \cline{2-6}
                                                                                & {\small FastText+Conv-KNRM~\cite{hofstatter2019effect} }& 0.277 & +14\% & 0.290 & +21\% \\
                                                                                & {\small BERT Re-Ranker~\cite{nogueira2019passage} }                                                          & 0.365                              & +50\%                             & 0.359                              & +50\% \\  \hline \hline
\end{tabular}
\endgroup

\end{table}

\footnotetext{Statistical significance is unknown because the MS MARCO website publishes only summary results.}

\textbf{First-stage search with DeepCT-Index vs. re-ranking.}  We further compared the \texttt{DeepCT-Index}-based BM25 retrieval to several competition systems by participating in the MS MARCO competition. 
Table~\ref{tab:leaderboard} shows results from the MS MARCO leaderboard\footnote{\url{http://www.msmarco.org/leaders.aspx}. Aug 13, 2019.}. It first lists 2 important first-stage rankers: the official standard BM25 and \texttt{Doc2Query}  \texttt{BM25}~\cite{nogueira2019doc2qry}. \texttt{DeepCT-Index} \texttt{BM25} outperformed both. Table~\ref{tab:leaderboard} also lists representative re-ranking approaches for feature-based learning-to-rank, previous state-of-the-art neural re-rankers (\emph{non-ensemble version}), and BERT-based re-rankers.  FastText+Conv-KNRM~\cite{hofstatter2019effect} was the best non-BERT model on the leaderboard. BERT Re-Ranker~\cite{nogueira2019passage} uses BERT as a black-box model that takes a query-passage pair and outputs a relevance scores; most recently-proposed re-rankers in the BERT family~\cite{nogueira2019passage, dai2019deeper} are based on this approach.  
All the multi-stage systems used the Official BM25 for first-stage ranking, and applied a re-ranker to fine-tune the rankings of the top 1000 passages.

\texttt{DeepCT-Index} \texttt{BM25} was better than several re-ranking pipelines. It is more accurate than feature-based learning-to-rank, a widely used re-ranking approach in modern search engines. It is also more accurate than a popular neural re-ranking model \texttt{K-NRM}~\cite{K-NRM}. Compared to \texttt{Duet V2} (the official re-ranking baseline) and \texttt{Conv-KNRM}~\cite{dai2018convolutional}, \texttt{DeepCT-Index BM25} achieves similar accuracy while being more efficient as it does not need the re-ranking stage. These results indicate that it is possible to replace some pipelined ranking systems with a single-stage retrieval using \texttt{DeepCT-Index}.

Finally, strong neural re-rankers like the BERT Re-Ranker were much more effective than \texttt{DeepCT-Index}  \texttt{BM25}. These models generate high quality soft-match signals between query and passage words (e.g. ``hotel'' to ``motel''). In contrast, \texttt{DeepCT-Index}  \texttt{BM25} only matches terms exactly, which provides much less evidence.  


\begin{table}[t]
\caption{Re-Ranking accuracy of two re-rankers applied to passages retrieved by BM25 from \texttt{DeepCT-Index} and the \texttt{tf} \texttt{index}. Dataset: MS MARCO dev.} \label{tab:rerank}

\begingroup
\setlength{\tabcolsep}{4pt}
\begin{tabular}{c|cc|cc|cc}
\hline \hline
      & \multicolumn{2}{c|}{Recall} & \multicolumn{2}{c|}{\begin{tabular}[c]{@{}c@{}}Conv-KNRM \\ Re-Ranker \\ MRR@10 \end{tabular}} & \multicolumn{2}{c}{\begin{tabular}[c]{@{}c@{}}BERT\\ Re-Ranker \\ MRR@10 \end{tabular}} \\

Depth &  \texttt{tf}        &  \texttt{DeepCT}        & \texttt{tf}                                     &  \texttt{DeepCT}                                  & \texttt{tf}                                   &  \texttt{DeepCT}                                \\ \hline

10    & 40\%         & 49\%            & 0.234                                   & 0.270                                   & 0.279                                & 0.320                                 \\
20    & 49\%         & 58\%            & 0.244                                   & 0.277                                   & 0.309                                & 0.343                                 \\
50    & 60\%         & 69\%            & 0.253                                   & 0.278                                   & 0.336                                & 0.361                                 \\
100   & 68\%         & 76\%            & 0.256                                   & 0.274                                   & 0.349                                & 0.368                                 \\
200   & 75\%         & 82\%            & 0.256                                   & 0.269                                   & 0.358                                & 0.370                                 \\
500   & 82\%         & 88\%            & 0.256                                   & 0.269                                   & 0.366                                & 0.374                                 \\
1000  & 86\%         & 91\%            & 0.256~$^{1}$                                   & 0.264                                   & 0.371~$^{1}$                                & 0.376                                
\\\hline \hline
\multicolumn{7}{p{8cm}}{$^{1}$\footnotesize{The values are not exactly the same as in Table 4 due to differences in the initial rankings generated from our BM25 and the official BM25 from MS MSARCO. }}
\end{tabular}
\endgroup
\end{table}

\footnotetext{We did not run this experiment on MS MARCO test set because test set results can only be evaluated by submitting to the MS MARCO competition. The organizers discourage too many submission from the same group to avoid "P-hacking". }

\textbf{\texttt{DeepCT-Index} as the first-stage of re-ranking systems}. 
The next experiment examines whether a first-stage ranking produced by \texttt{DeepCT-Index} \texttt{BM25} can improve later-stage re-rankers. Table~\ref{tab:rerank} reports the performance of two re-rankers applied to candidate passages retrieved from \texttt{DeepCT-Index} and the \texttt{tf} index.
The re-rankers were selected based on their performance on the MSMARCO leaderboard (Table 4): Conv-KNRM~\cite{dai2018convolutional}, which has medium accuracy, and BERT Re-Ranker~\cite{nogueira2019passage}, which has high accuracy. We tested various re-ranking depths. Re-ranking at a shallower depth has higher efficiency but may miss more relevant passages. 

The recall values show the percentage of relevant passages in the re-ranking passage set. 
\texttt{DeepCT-Index} had higher recall at all depths, meaning a ranking from  \texttt{DeepCT-Index} provided more relevant passages to a re-ranker. Both re-rankers consistently achieved higher MRR@10 by using \texttt{DeepCT-Index} compared to using \texttt{tf} index. For Conv-KNRM, the best MRR@10 improved from $0.256$ to $0.278$, and the required re-ranking depth decreased from $100$ to $50$. In other words,  a first-stage ranking from \texttt{DeepCT-Index} helped the re-ranker to be over $8\%$ more accurate and $2\times$ more efficient. 
For BERT re-ranker, 
\texttt{DeepCT-Index} enabled the re-ranker to achieve similar accuracy using much fewer passages. Re-ranking the top 100-200 passages from \texttt{DeepCT-Index} produced similar MRR@10 as re-ranking the top 1,000 passages from \texttt{tf} index.  

The high computational cost of deep neural-based re-rankers is one of the biggest concerns about adopting them in online services. ~\citet{nogueira2019doc2qry} reported that adding a BERT Re-Ranker, with a re-ranking depth of 1000, introduces $10\times$ more latency to a BM25 first-stage ranking even using GPUs or TPUs. 
\texttt{DeepCT-Index} reduces the re-ranking depth by $5\times$ to $10\times$,
making deep neural-based re-rankers practical in latency-/resource-sensitive systems.

\textbf{Summary.} There is much prior research about passage term weighting, but it has not been clear how to effectively model a word's syntax and semantics in specific passages. Our results show that a deep, contextualized neural language model is able to capture some of the desired properties, and can be used to generate  effective term weights for passage indexing. A BM25 retrieval on \texttt{DeepCT-Index} can be $25\%$ more accurate than classic \textit{tf}-based indexes, and are more accurate than some widely-used multi-stage retrieval systems. The improved first stage ranking further benefits the effectivenss and efficiency of downstream re-rankers. 

\begin{table*}[t]
\caption{Visualization of \texttt{DeepCT-Index} passage term weights. Red shades reflect the normalized term weights -- the percentage of total passage term weights applied to the term. White indicates zero weight. Query terms are bold.  } \label{tab:weight_viz}
\small

\begin{tabular}{p{0.07\textwidth} p{0.87\textwidth} }

\hline \hline
& Percentage of weights a term takes in the passage: \begin{CJK}{UTF8}{gbsn}{\setlength{\fboxsep}{0pt}\colorbox{white!0}{\parbox{0.8\textwidth}{
\colorbox{red!0}{\strut  0} 
\colorbox{red!20}{\strut $10\%$}
\colorbox{red!40}{\strut $20\%$}
\colorbox{red!60}{\strut $30\%$}
\colorbox{red!80}{\strut $40\%$}
\colorbox{red!100}{\strut >$50\%$ }
}}}\end{CJK} \\ \hline
Query & \textbf{who is susan boyle} \\
On-Topic & \begin{CJK*}{UTF8}{gbsn}
{\setlength{\fboxsep}{0pt}\colorbox{white!0}{\parbox{0.87\textwidth}{
\colorbox{red!4}{\strut Amateur} \colorbox{red!5}{\strut vocalist} \colorbox{red!83}{\strut \textbf{Susan}} \colorbox{red!85}{\strut \textbf{Boyle}} \colorbox{red!0}{\strut became} \colorbox{red!0}{\strut an} \colorbox{red!0}{\strut overnight} \colorbox{red!1}{\strut sensation} \colorbox{red!0}{\strut after} \colorbox{red!2}{\strut appearing} \colorbox{red!0}{\strut on} \colorbox{red!0}{\strut the} \colorbox{red!0}{\strut first} \colorbox{red!0}{\strut round} \colorbox{red!0}{\strut of} \colorbox{red!0}{\strut 2009}\colorbox{red!0}{\strut 's} \colorbox{red!2}{\strut popular} \colorbox{red!0}{\strut U.K.} \colorbox{red!3}{\strut reality} \colorbox{red!7}{\strut show} \colorbox{red!0}{\strut Britain}\colorbox{red!0}{\strut 's} \colorbox{red!4}{\strut Got} \colorbox{red!4}{\strut Talent}\colorbox{red!0}{\strut .} 
}}}
\end{CJK*} \\
Off-Topic & 
\begin{CJK*}{UTF8}{gbsn}
{\setlength{\fboxsep}{0pt}\colorbox{white!0}{\parbox{0.87\textwidth}{
\colorbox{red!0}{\strut Best} \colorbox{red!0}{\strut Answer}\colorbox{red!0}{\strut :} \colorbox{red!0}{\strut a} \colorbox{red!38}{\strut troll} \colorbox{red!0}{\strut is} \colorbox{red!3}{\strut generally} \colorbox{red!0}{\strut someone} \colorbox{red!0}{\strut who} \colorbox{red!0}{\strut tries} \colorbox{red!0}{\strut to} \colorbox{red!0}{\strut get} \colorbox{red!2}{\strut attention} \colorbox{red!0}{\strut by} \colorbox{red!0}{\strut posting} \colorbox{red!0}{\strut things} \colorbox{red!0}{\strut everyone} \colorbox{red!0}{\strut will} \colorbox{red!0}{\strut disagree}\colorbox{red!0}{\strut ,} \colorbox{red!0}{\strut like} \colorbox{red!0}{\strut going} \colorbox{red!0}{\strut to} \colorbox{red!0}{\strut a} \colorbox{red!19}{\strut \textbf{susan}} \colorbox{red!3}{\strut \textbf{boyle}} \colorbox{red!0}{\strut fan} \colorbox{red!0}{\strut page} \colorbox{red!0}{\strut and} \colorbox{red!2}{\strut writing} \colorbox{red!19}{\strut \textbf{susan}} \colorbox{red!3}{\strut \textbf{boyle}} \colorbox{red!0}{\strut is} \colorbox{red!0}{\strut ugly} \colorbox{red!0}{\strut on} \colorbox{red!0}{\strut the} \colorbox{red!0}{\strut wall}\colorbox{red!0}{\strut .} \colorbox{red!0}{\strut they} \colorbox{red!0}{\strut are} \colorbox{red!3}{\strut usually} \colorbox{red!0}{\strut 14-16} \colorbox{red!0}{\strut year} \colorbox{red!5}{\strut olds} \colorbox{red!0}{\strut who} \colorbox{red!3}{\strut crave} \colorbox{red!0}{\strut attention.}
}}}
\end{CJK*} \\ \hline

Query & \textbf{what values do zoos serve} \\
On-Topic & \begin{CJK}{UTF8}{gbsn}
{\setlength{\fboxsep}{0pt}\colorbox{white!0}{\parbox{0.87\textwidth}{
\colorbox{red!54}{\strut \textbf{Zoos}} \colorbox{red!9}{\strut \textbf{serve}} \colorbox{red!0}{\strut several} \colorbox{red!58}{\strut purposes} \colorbox{red!0}{\strut depending} \colorbox{red!0}{\strut on} \colorbox{red!0}{\strut who} \colorbox{red!0}{\strut you} \colorbox{red!0}{\strut ask}\colorbox{red!0}{\strut .} \colorbox{red!0}{\strut 1}\colorbox{red!0}{\strut )} \colorbox{red!0}{\strut Park/Garden}\colorbox{red!0}{\strut :} \colorbox{red!0}{\strut Some} \colorbox{red!54}{\strut \textbf{zoos}} \colorbox{red!0}{\strut are} \colorbox{red!1}{\strut similar} \colorbox{red!0}{\strut to} \colorbox{red!0}{\strut a} \colorbox{red!2}{\strut botanical} \colorbox{red!6}{\strut garden} \colorbox{red!0}{\strut or} \colorbox{red!5}{\strut city} \colorbox{red!3}{\strut park}\colorbox{red!0}{\strut .} \colorbox{red!0}{\strut They} \colorbox{red!0}{\strut give} \colorbox{red!1}{\strut people} \colorbox{red!2}{\strut living} \colorbox{red!0}{\strut in} \colorbox{red!0}{\strut crowded}\colorbox{red!0}{\strut ,} \colorbox{red!0}{\strut noisy} \colorbox{red!3}{\strut cities} \colorbox{red!0}{\strut a} \colorbox{red!0}{\strut place} \colorbox{red!0}{\strut to} \colorbox{red!0}{\strut walk} \colorbox{red!0}{\strut through} \colorbox{red!0}{\strut a} \colorbox{red!0}{\strut beautiful}\colorbox{red!0}{\strut ,} \colorbox{red!0}{\strut well} \colorbox{red!0}{\strut maintained} \colorbox{red!2}{\strut outdoor} \colorbox{red!0}{\strut area}\colorbox{red!0}{\strut .} \colorbox{red!0}{\strut The} \colorbox{red!7}{\strut animal} \colorbox{red!1}{\strut exhibits} \colorbox{red!0}{\strut create} \colorbox{red!0}{\strut interesting} \colorbox{red!0}{\strut scenery} \colorbox{red!0}{\strut and} \colorbox{red!0}{\strut make} \colorbox{red!0}{\strut for} \colorbox{red!0}{\strut a} \colorbox{red!0}{\strut fun} \colorbox{red!0}{\strut excursion}\colorbox{red!0}{\strut .} 
}}}
\end{CJK}
\\
Off-topic & \begin{CJK}{UTF8}{gbsn}
{\setlength{\fboxsep}{0pt}\colorbox{white!0}{\parbox{0.87\textwidth}{
\colorbox{red!0}{\strut There} \colorbox{red!0}{\strut are} \colorbox{red!0}{\strut NO} \colorbox{red!12}{\strut purebred} \colorbox{red!17}{\strut Bengal} \colorbox{red!34}{\strut tigers} \colorbox{red!0}{\strut in} \colorbox{red!0}{\strut the} \colorbox{red!0}{\strut U.S}\colorbox{red!0}{\strut .} \colorbox{red!0}{\strut The} \colorbox{red!0}{\strut only} \colorbox{red!12}{\strut purebred} \colorbox{red!34}{\strut tigers} \colorbox{red!0}{\strut in} \colorbox{red!0}{\strut the} \colorbox{red!0}{\strut U.S.} \colorbox{red!0}{\strut are} \colorbox{red!0}{\strut in} \colorbox{red!2}{\strut AZA} \colorbox{red!4}{\strut \textbf{zoos}} \colorbox{red!0}{\strut and} \colorbox{red!0}{\strut include} \colorbox{red!0}{\strut 133} \colorbox{red!2}{\strut Amur} \colorbox{red!0}{\strut (}\colorbox{red!0}{\strut AKA} \colorbox{red!2}{\strut Siberian}\colorbox{red!0}{\strut )}\colorbox{red!0}{\strut ,} \colorbox{red!0}{\strut 73} \colorbox{red!0}{\strut Sumatran} \colorbox{red!0}{\strut and} \colorbox{red!0}{\strut 50} \colorbox{red!0}{\strut Malayan} \colorbox{red!34}{\strut tigers} \colorbox{red!0}{\strut in} \colorbox{red!0}{\strut the} \colorbox{red!3}{\strut Species} \colorbox{red!0}{\strut Survival} \colorbox{red!0}{\strut Plan}\colorbox{red!0}{\strut .} \colorbox{red!0}{\strut All} \colorbox{red!0}{\strut other} \colorbox{red!0}{\strut U.S.} \colorbox{red!2}{\strut captive} \colorbox{red!34}{\strut tigers} \colorbox{red!0}{\strut are} \colorbox{red!5}{\strut inbred} \colorbox{red!0}{\strut and} \colorbox{red!0}{\strut cross} \colorbox{red!2}{\strut bred} \colorbox{red!0}{\strut and} \colorbox{red!0}{\strut do} \colorbox{red!0}{\strut not} \colorbox{red!0}{\strut \textbf{serve}} \colorbox{red!0}{\strut any} \colorbox{red!0}{\strut conservation} \colorbox{red!0}{\strut \textbf{value}}\colorbox{red!0}{\strut .} 
}}}
\end{CJK} \\ \hline

Query & \textbf{do atoms make up dna} \\ 

On-Topic & \begin{CJK*}{UTF8}{gbsn}
{\setlength{\fboxsep}{0pt}\colorbox{white!0}{\parbox{0.87\textwidth}{
\colorbox{red!23}{\strut \textbf{DNA}} \colorbox{red!0}{\strut only} \colorbox{red!0}{\strut has} \colorbox{red!0}{\strut 5} \colorbox{red!4}{\strut different} \colorbox{red!22}{\strut \textbf{atoms}} \colorbox{red!0}{\strut -} \colorbox{red!2}{\strut carbon}\colorbox{red!0}{\strut ,} \colorbox{red!1}{\strut hydrogen}\colorbox{red!0}{\strut ,} \colorbox{red!1}{\strut oxygen}\colorbox{red!0}{\strut ,} \colorbox{red!1}{\strut nitrogen} \colorbox{red!0}{\strut and} \colorbox{red!1}{\strut phosphorous}\colorbox{red!0}{\strut .} \colorbox{red!0}{\strut According} \colorbox{red!0}{\strut to} \colorbox{red!0}{\strut one} \colorbox{red!0}{\strut estimation}\colorbox{red!0}{\strut ,} \colorbox{red!0}{\strut there} \colorbox{red!0}{\strut are} \colorbox{red!0}{\strut about} \colorbox{red!0}{\strut 204} \colorbox{red!0}{\strut billion} \colorbox{red!22}{\strut \textbf{atoms}} \colorbox{red!0}{\strut in} \colorbox{red!0}{\strut each} \colorbox{red!23}{\strut \textbf{DNA}}\colorbox{red!0}{\strut .} 
}}}
\end{CJK*} \\

Off-Topic & \begin{CJK*}{UTF8}{gbsn}
{\setlength{\fboxsep}{0pt}\colorbox{white!0}{\parbox{0.87\textwidth}{
\colorbox{red!46}{\strut Genomics} \colorbox{red!0}{\strut in} \colorbox{red!10}{\strut Theory} \colorbox{red!0}{\strut and} \colorbox{red!6}{\strut Practice}\colorbox{red!0}{\strut .} \colorbox{red!0}{\strut What} \colorbox{red!0}{\strut is} \colorbox{red!46}{\strut Genomics}\colorbox{red!0}{\strut .} \colorbox{red!46}{\strut Genomics} \colorbox{red!0}{\strut is} \colorbox{red!0}{\strut a} \colorbox{red!5}{\strut study} \colorbox{red!0}{\strut of} \colorbox{red!0}{\strut the} \colorbox{red!4}{\strut genomes} \colorbox{red!0}{\strut of} \colorbox{red!4}{\strut organisms}\colorbox{red!0}{\strut .} \colorbox{red!0}{\strut It} \colorbox{red!3}{\strut main} \colorbox{red!0}{\strut task} \colorbox{red!0}{\strut is} \colorbox{red!0}{\strut to} \colorbox{red!4}{\strut determine} \colorbox{red!0}{\strut the} \colorbox{red!0}{\strut entire} \colorbox{red!3}{\strut sequence} \colorbox{red!0}{\strut of} \colorbox{red!6}{\strut \textbf{DNA}} \colorbox{red!0}{\strut or} \colorbox{red!0}{\strut the} \colorbox{red!0}{\strut composition} \colorbox{red!0}{\strut of} \colorbox{red!0}{\strut the} \colorbox{red!3}{\strut \textbf{atoms}} \colorbox{red!0}{\strut that} \colorbox{red!0}{\strut make} \colorbox{red!0}{\strut up} \colorbox{red!0}{\strut the} \colorbox{red!6}{\strut \textbf{DNA}} \colorbox{red!0}{\strut and} \colorbox{red!0}{\strut the} \colorbox{red!0}{\strut chemical} \colorbox{red!0}{\strut bonds} \colorbox{red!0}{\strut between} \colorbox{red!0}{\strut the} \colorbox{red!6}{\strut \textbf{DNA}} \colorbox{red!3}{\strut \textbf{atoms}}\colorbox{red!0}{\strut .} 
}}}
\end{CJK*}\\ \hline \hline

\end{tabular}
\end{table*}

\subsection{Understanding Sources of Effectiveness}

This section aims to understand the sources of effectiveness of \texttt{DeepCT-Index} through several analyses.

Table~\ref{tab:weight_viz} visualizes \texttt{DeepCT-Index} weights on a few cases. Each case has a query, a relevant passage, and a non-relevant passage that mentions query concepts but is actually off-topic. The heatmap visualizes the term weights in \texttt{DeepCT-Index}. Weights are normalized by the sum of all term weights in the passage, to reflect relative term importance in the passage.

\textbf{Emphasize central terms and suppress non-central terms}. \texttt{DeepCT} is able to identify terms that are central to the topic of the text.  In the first query in Table~\ref{tab:weight_viz}, both passages mention \textit{``susan boyle''}. In the relevant passage, \texttt{DeepCT-Index} recognized that the topic is \textit{``susan boyle''} and put almost all weight on these two terms. The off-topic passage is about the definition of \textit{``troll''}, while \textit{``susan boyle''} is used in an example.  \texttt{DeepCT-Index} managed to identify the central concept \textit{``troll''} and suppress other non-topical terms; less than $10\%$ of weight goes to \textit{``susan boyle''}. With the new weights, the BM25 score between the query and the off-topic passage is greatly decreased.  Similar behavior can be seen on other cases in Table~\ref{tab:weight_viz}. 

\begin{figure}[tb]
\centering
\begin{subfigure}[b]{0.21\textwidth}
\includegraphics[width=\textwidth]{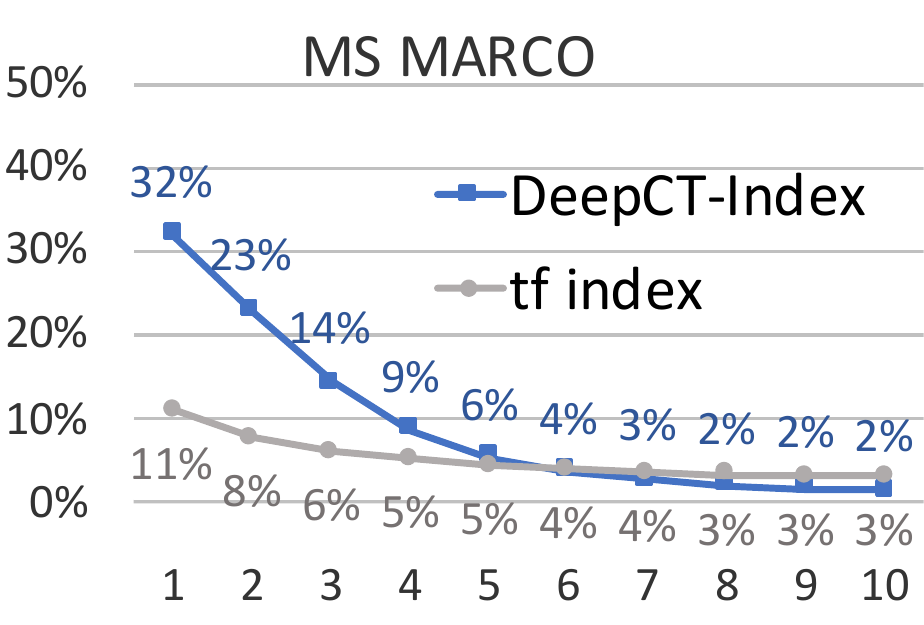}
\end{subfigure}%
\begin{subfigure}[b]{0.21\textwidth}
\includegraphics[width=\textwidth]{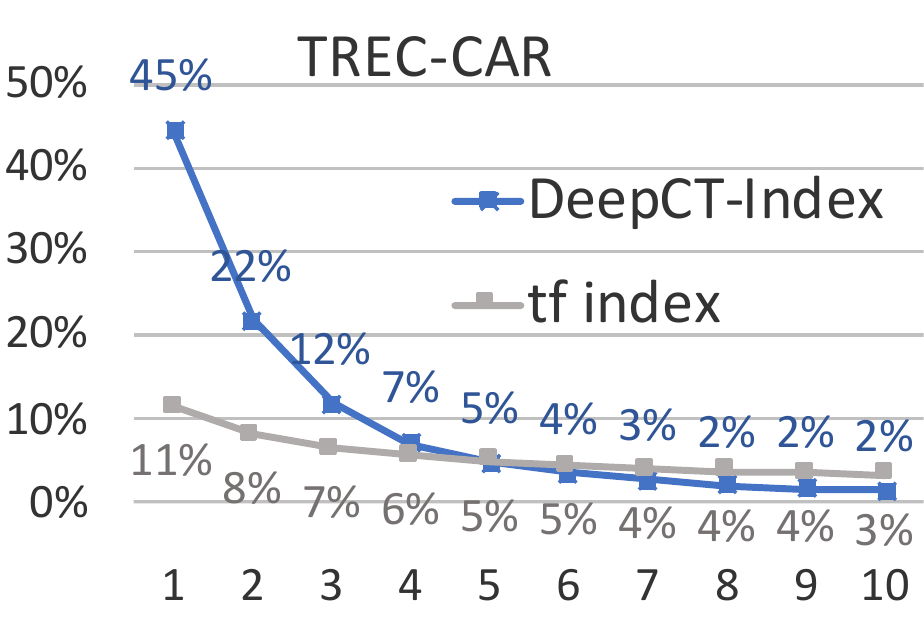}
\end{subfigure}
\caption{Term weight distribution of the top-10 terms in passages with highest weights. The X-axis shows the term's rank ordered by weight. The Y-axis shows the average term weight normalized by total passage term weight.}~\label{fig:term_dist}
\label{fig:DeepCT}
\end{figure}

Figure~\ref{fig:DeepCT} compares the term weight distribution of \texttt{DeepCT-Index} and \texttt{tf} index. It plots the average weight of each passage's highest-weighted term, the average weight of the second highest-weighted term, and so on. The original \textit{tf} distribution is flat. \texttt{DeepCT-Index} assigns high weights to a few central terms, resulting in a skewed term weight distribution. Such skewed distribution confirms our observations from the case study that \texttt{DeepCT-Index} aggressively emphasizes a few central terms and supresses the others.
 
\texttt{DeepCT}'s strong bias towards central terms explains its effectiveness: promoting on-topic terms and suppressing off-topic terms. It is also a source of \emph{error}. In many failing cases, \texttt{DeepCT-Index} identified the central passage words correctly, but the answer to the query is mentioned in a non-central part of the passage.  \texttt{DeepCT-Index} down-weighted or even ignored these parts, causing relevant passage to be ranked lower. It is worth exploring how to reduce the risk of losing information that is useful but not central.

\textbf{Context-based weights vs. frequency-based weights}. The examples in Table~\ref{tab:weight_viz} also show that the term weights produced by \texttt{DeepCT-Index} are based on the meaning of the context rather than frequency.  A term may get low weight even it is frequent (e.g. in the last \textit{``Genomics''} passage, \textit{``DNA''} is considered unimportant even though it is mentioned 3 times ). The same term receives very different weights in different passage even when \textit{tf} is the same. This extent of independence from frequency signals is uncommon in previous term weighting approaches; even semantic text representations such as Word2Vec~\cite{word2vec} and Glove~\cite{pennington2014glove} are reported to have high correlation with term frequency 
~\cite{schnabel2015evaluation}.

\section{Experimental Methodology and Results for D\texorpdfstring{\lowercase{eep}}{deep}CT-Q\texorpdfstring{\lowercase{uery}}{uery}}
\label{section:exp-set-qry}

\begin{table*}[t]
\caption{Retrieval accuracy of \texttt{DeepCT-Query} using \texttt{QL} retrieval model on Robust04 and Gov2. $^*$ indicates statistically significant improvements of \texttt{BOW-DeepCT-Query} over \texttt{BOW-DeepTR}. $^\dagger$ indicates statistically significant improvements of \texttt{SDM-DeepCT-Query} over \texttt{SDM-DeepTR}. \texttt{BOW}-/\texttt{SDM}-\textit{Oracle} weight terms by ground truth, estimating an upper limit for \texttt{DeepTR} and \texttt{DeepCT-Query}. }\label{tab:query}

\begingroup
\setlength{\tabcolsep}{3pt}
\renewcommand{\arraystretch}{0.9}
\begin{tabular}{l|cccccc||cccccc}
\hline\hline
             & \multicolumn{6}{c||}{Robust04}                        & \multicolumn{6}{c}{Gov2}                            \\ \cline{2-13}
   \multirow{2}{*}{Query}          &
   \multicolumn{2}{c}{Title} &
   \multicolumn{2}{c}{Description} & \multicolumn{2}{c||}{Narrative} &
   \multicolumn{2}{c}{Title} &
   \multicolumn{2}{c}{Description} & \multicolumn{2}{c}{Narrative} \\
             & {\small NDCG@20}      &{\small MAP}   & {\small NDCG@20}      & {\small MAP}     & {\small NDCG@20}      & {\small MAP}        & {\small NDCG@20}      & {\small MAP} & {\small NDCG@20}      & {\small MAP}       & {\small NDCG@20}      & {\small MAP}         \\ \hline

{\small\texttt{BOW}-\texttt{tf} }      & \textbf{0.410} &  \textbf{0.243}  & 0.417        & 0.248     & 0.320        & 0.177    & 0.442  & 0.291 & 0.407        & 0.253     & 0.419        & 0.231     \\

{\small\texttt{BOW-DeepTR} }   & 0.381 & 0.226   & 0.427        & 0.264     & 0.330        & 0.183     &0.437 & 0.289 &   0.427     & 0.271     & 0.395        & 0.211     \\ 
{\small\texttt{BOW-DeepCT-Query}}  & 0.383 & 0.229 & \textbf{0.445}$^*$        & \textbf{0.273}$^*$     & \textbf{0.367}$^*$         & \textbf{0.213}$^*$  & \textbf{0.443} &   \textbf{0.293} &  \textbf{0.430}        & \textbf{0.280}     & \textbf{0.438}$^*$         & \textbf{0.260}$^*$  \\  
{\small\texttt{BOW}-\textit{Oracle} } & \textit{0.369} & \textit{0.228} &   \textit{0.484}       &  \textit{0.311}       & \textit{0.447}  & \textit{0.283}  &  \textit{0.453}  &    \textit{0.306}     &  \textit{0.462}    &  \textit{0.312} &     \textit{0.481}   &    \textit{0.298} 
\\ \hline \hline
{\small\texttt{SDM-tf} }      & \textbf{0.427} & \textbf{0.264}   & 0.427        & 0.261     & 0.332        & 0.186   & \textbf{0.483} &  \textbf{0.324} &    0.434          & 0.270          &      0.436       &     0.239      \\ 

{\small\texttt{SDM-DeepTR}}   & 0.394 &  0.241   & 0.452        & 0.261     & 0.355        & 0.186    & 0.482 & 0.324 &     \textbf{0.464}   &  \textbf{0.294}    &     0.432    &  0.237    \\
{\small\texttt{SDM-DeepCT-Query}}  & 0.394 & 0.242 & \textbf{0.462}        & \textbf{0.288}$^\dagger$     & \textbf{0.380}$^\dagger$        & \textbf{0.225}$^\dagger$ & 0.483 & 0.323   &  0.446       & 0.292    &   \textbf{0.455}$^\dagger$      &  \textbf{0.268}$^\dagger$
\\
{\small\texttt{SDM}-\textit{Oracle}}  & \textit{0.398} & \textit{0.248} &  \textit{0.453}      & \textit{0.295}       &  \textit{0.441} & \textit{0.276}  & \textit{0.492}  &  \textit{0.337}       &  \textit{0.472}   &  \textit{0.319}  &   \textit{0.496}    & \textit{0.305}   
\\ 
\hline\hline
\end{tabular}
\endgroup
\end{table*}

This section presents the experimental methodology and results for the query term weighting task.  
The methodolgy follows the settings used in~\cite{zheng2015learning}. 

\textbf{Datasets:} We used the \textbf{Robust04} and \textbf{Gov2} TREC collections. Robust04 is a news corpus with 0.5M documents
and 249 test topics.  Gov2 is a web collection with 25M web pages and 150 test topics. Each test topic has 3 types of query: a short title query, a sentence-long query
description, and a passage-long query narrative. 

\textbf{Baselines}: Experiments were done with 6 baselines that use two forms of query structure (\texttt{BOW}, \texttt{SDM}) and three types of term weights (\texttt{tf}, \texttt{DeepTR}, \textit{Oracle}). \texttt{BOW} and \texttt{SDM} stand for bag-of-word queries and sequential dependency queries~\cite{metzler2005markov}. \texttt{tf} is the classic query term frequency weights. 
\texttt{DeepTR} is a previous state-of-the-art query weighting approach~\cite{zheng2015learning}.  
It extracts word features using the difference between the word's own embedding and the average embedding of the query. The features are linearly combined to produce term weights.\texttt{DeepTR} is supervised trained on term recall (Eq. 3). \textit{Oracle} weights query terms by the ground truth term recall weights; it reflects how much \texttt{DeepTR} and \texttt{DeepCT-Query} can achieve if they made perfect predictions, estimating an upper limit.


\textbf{Indexing, Retrieval and Evaluation}: Following~\citet{zheng2015learning}, we use the Indri search engine\footnote{\url{http://lemurproject.org/indri/}} with standard stemming and stop words filtering. We train and evaluate \texttt{DeepCT-Query} and \texttt{DeepTR} with 5-fold cross validation. ~\citet{zheng2015learning} show that query likelihood (\texttt{QL}) model performs slightly better than BM25, so we use \texttt{QL} to search the index. Retrieval results are evaluated using standard TREC metrics: NDCG@20 and MAP@1000.

\textbf{DeepCT Settings}: 
The BERT part of \texttt{DeepCT} was initialized with the official pre-trained BERT (uncased, base model)~\cite{devlin2018bert}. \texttt{DeepCT}, including the BERT layers and the last regression layer, was fine-tuned end-to-end. The model was trained for $10$ epochs. We used a learning rate of $2e^{-5}$. Max input text length was set to be $30$, $50$ and $100$ tokens for query titles, descriptions, and narratives.

\textbf{Results}. Results are listed in Table ~\ref{tab:query}. 
The short title queries did not benefit from the term weighting approaches. Title queries often consist of a few keywords that are all essential, so re-weighting is less important. Besides, there isn't much context for \texttt{DeepCT} to leverage to estimate term importance. External information about the query, such as pseudo-relevance feedback~\cite{lavrenko2001relevance}, may be necessary to understand short queries. 

Description and narrative queries mention many terms and concepts; it is important to identify which are central during retrieval. Weighted queries are more effective than the un-weighted queries. \texttt{DeepCT-Query} is more accurate than \texttt{DeepTR} in most cases. \texttt{DeepCT-Query} differs from \texttt{DeepTR} in how they represent a term with respect to the context. The results demonstrate that \texttt{DeepCT-Query} better reflect a word's role in the query. Larger improvements were observed on narrative queries than on description queries. 
The results indicate that for short sentences, simple context modeling may be effective. But for more complex queries, a deep language modeling component like BERT can lead to improved search results. 




\section{Conclusion}\label{section:conclusion}

Recently much research has focused on the later stages of multi-stage search engines.  
Most first-stage rankers are older-but-efficient bag-of-words retrieval models that use term frequency signals. However, frequency-based term weighting does not necessarily reflect a term's importance in queries and documents, especially when the frequency distribution is flat, such as in sentence-long queries or passage-long documents. More accurate term importance estimates require the system to understand the role each word plays in each specific context. This paper presents \texttt{DeepCT}, a novel context-aware term weighting approach that better estimates term importance for first-stage bag-of-words retrieval systems.

The \texttt{DeepCT} framework is built on BERT~\cite{devlin2018bert}, a recent deep neural language model.
In \texttt{BERT}, a word's embedding gradually `absorbs' information from other related words in the input text and generates a new word embedding that characterizes the word in a specific context. \texttt{DeepCT} uses BERT to extract context-based term features and learns to use these features to predict the importance for each term in a supervised per-token regression task. The training signals are mined from relevance-based query-document pairs so that the predicted term weights are aligned with the retrieval task.

One use of \texttt{DeepCT} is \texttt{DeepCT-Index}, which weights document terms.  \texttt{DeepCT} produces integer term weights that can be stored in a typical inverted index and are compatible with popular first-stage retrieval models such as BM25 and query likelihood. Another use of \texttt{DeepCT} is \texttt{DeepCT-Query}, which weights query terms. Experimental results show that \texttt{DeepCT-Query} greatly improves the accuracy of longer queries, due to its ability to identify central query terms in a complex context.

Experimental results  show that \texttt{DeepCT-Index} improves the accuracy of two popular first-stage retrieval algorithms by up to $50\%$.  Running \texttt{BM25} on \texttt{DeepCT-Index} can be as effective as several previous state-of-the-art multi-stage search systems that use knowledge bases, machine learning, and large amounts of training data.

The higher-quality ranking enabled by \texttt{DeepCT-Index} improves the accuracy/efficiency tradeoff for later-stage re-rankers. A state-of-the-art BERT-based re-ranker achieved similar accuracy with $5\times$ fewer candidate documents, making such computation-intensive re-rankers more practical in latency-/resource-sensitive systems. 


Although much progress has been made toward developing better neural ranking models for IR, computational complexity often limits these models to the re-ranking stage. \texttt{DeepCT} successfully transfers the text understanding ability from a deep neural network into simple signals that can be efficiently consumed by early-stage ranking systems and boost their performance. 

Analysis shows the main advantage of \texttt{DeepCT} over classic term-weighting approaches: \texttt{DeepCT} finds the most central words in a text even if they are mentioned only once. Non-central words, even if mentioned frequently in the text, are suppressed. Such behavior is uncommon in previous term weighting approaches. We view \texttt{DeepCT} as an encouraging step from ``frequencies'' to `` meanings''.


\clearpage

\bibliographystyle{ACM-Reference-Format}
\bibliography{bibliography}



\end{document}